\documentclass{elsart}
\usepackage{amssymb}
\usepackage{graphicx,amssymb,lineno}

\begin{document}
\begin{frontmatter}
\title{Binding Energies and Radii of the Nuclei with $N\geq Z$  in an Alpha-Cluster Model}

\author{G. K. NIE}

\address{Institute of Nuclear Physics, Ulugbek, Tashkent 702132, Uzbekistan}
\ead{galani@Uzsci.net}
\begin{abstract}
Using the surface tension energy put in dependence on the number of
$\alpha$-clusters in the core in a phenomenological model representing a nucleus as
a core and a nuclear molecule on its surface leads to widening the number of
isotopes to be described from the narrow strip of $\beta$-stability to the isotopes
with $N \geq Z$. The number of alpha-clusters in the molecule is obtained from the
analysis of experimental binding energies and the specific density of the core
binding energy $\rho $ and the radii are calculated. It is shown that for the
isotopes of one $Z$ with growing $A$ the number of alpha-clusters of the molecule
decreases mostly to 3 and $\rho$ increases to reach a saturation value within $\rho
= 2.5 \div 2.7$ $\rm MeV/fm^3$ at the $\beta$ - stable isotopes, so the narrow strip
of the binding energies of the $\beta$-stable isotopes with $Z\leq$ 84 is outlined
by a function of one variable $Z$.
\end{abstract}
\end{frontmatter}

\section{The formulas to calculate binding energies and radii}

In the representation of a nucleus as a core and a nuclear molecule on the surface
of the core the nuclear binding energies and the radii of $\beta$- stable isotopes
are calculated \cite{1}. The calculations are made in the framework of an $\alpha
$-cluster model based on pn-pair interactions with using isospin invariance of
nuclear force \cite{2,3}. In the framework of the model some parameters like the
binding energy and the energy of Coulomb repulsion between two nearby clusters had
been found to be $\epsilon _{\alpha \alpha }=2.425$ MeV and $\epsilon _{\alpha
\alpha }^{C}=1.925$ MeV, so that the nuclear force energy is to be $\epsilon
_{\alpha \alpha }^{nuc}=\epsilon _{\alpha \alpha }+\epsilon _{\alpha \alpha
}^{C}=4.350$ MeV \cite{3} as well as the energy of nuclear force of one $\alpha
$-cluster $ \epsilon _{\alpha }^{nuc}=\epsilon _{\alpha }+\epsilon _{\alpha
}^{C}=29.060$ MeV \cite{3,4} where $\epsilon _{\alpha }=E_{^{4}\mathrm{He}}=28.296$
MeV and $\epsilon _{\alpha }^{C}=0.764$ MeV.

One of the main findings \cite{3} of the phenomenological model is the formula for
binding energy $E$ of the symmetrical nuclei $(N=Z,Z < 30)$ with the number of
$\alpha$ - clusters $N_\alpha= Z/2$

\begin{equation}
E=N_{\alpha }\epsilon _{\alpha }+3(N_{\alpha }-2)\epsilon _{\alpha \alpha },
\label{1}
\end{equation}
for the odd $Z_{1}=Z+1$ nuclei the energy $E_{1}=E+13.9$ MeV. It is suggested that
the number of short range links in a nucleus consisting of $ N_{\alpha }$ $\alpha
$-clusters is $3(N_{\alpha }-2)$ and that the long range part of the Coulomb
interactions must be compensated with the surface tension energy $E^{st}$. Then with
using isospin invariance of nuclear force the empirical values of the Coulomb
energy, the energy of surface tension and the empirical values of the distance of
the position of the last alpha-cluster $R_{\alpha }$ in the system of the center of
the masses of the remote $N_{\alpha }-4$ $\alpha $-clusters have been obtained
\cite{3}. From the analysis of these values the formulas for Coulomb radius $
R_{C}=1.869N_{\alpha }^{1/3}$fm (see (22) \cite{3}), for the radius of the position
of the last $\alpha $-cluster $R_{\alpha }=2.168(N_{\alpha }-4)^{1/3}$ fm (see (21)
\cite{2,3}) have been found. The Coulomb energy of the charge sphere of the radius
$R_{C}$ is $E^{C}=3/5Z^{2}e^{2}/R_{C}$, which after simplifying is $1.848(N_{\alpha
})^{5/3}$ MeV \cite{3}. The binding energy of the excess nn-pairs for the
beta-stable nuclei $E_{\Delta N}=\sum_{1}^{N_{nn}}E_{i_{nn}}$ (see (13 ) \cite{3})
 where $E_{i_{nn}}$ is the binding energy of the $i^{th}$ nn-pair in the core. The
nn-pairs fill out the free space in the core which appears due to the difference
between the charge and the matter radii of an alpha-cluster. Thus, the binding
energy of all $\beta $ - stable nuclei with an accuracy in a few MeV is calculated
as the sum $E=E^{nuc}+E^{st}-E^{C}+E_{\Delta N}$ where $E^{nuc}$ is the energy of
nuclear force in short range links (see (7) \cite{3}).

There is a clear relation between the nuclei $A(Z, N)$ and $A_1(Z_1, N+2)$ where
$Z_1=Z+1$, $N$ is an even number and $N\geq Z$ \cite{1,2,3}. Then $A_1 = A + 3$. The
nuclei have equal cores with $N_\alpha^{core}$ $\alpha$ - clusters and with the same
number of excess nn-pairs $ N_{nn} = \Delta N/2$, where $\Delta N = N - Z$. The
peripheral molecule on the surface of the core consists of $N_\alpha^{ml}$ $\alpha$
- clusters for the even nucleus and of $N_\alpha^{ml}$ + 0.5 for the odd nucleus. In
case of the nucleus $ A_1(Z_1, N+2)$ one neutron is stuck to the single pn-pair. It
has been taken into account \cite{1} that the number of short range links in the
core has to be $3(N_\alpha^{core} - 2 )$ + 6, because the total number of links is $
3(N_\alpha - 2) = 3(N_\alpha^{ml} - 2) + 3 (N_\alpha^{core} - 2 )+6$. Therefore the
energy of six short range links $ 6\epsilon_{\alpha \alpha} = 6(\epsilon_{\alpha
\alpha}^{nuc} - \epsilon_{\alpha \alpha}^{C})$ is added to the binding energy of
$N_\alpha^{core}$ $\alpha$ - clusters in calculation of the core binding energy.

So the binding energies $E$ and $E_{1}$ of the nuclei $A(Z,N)$ and
$A_{1}(Z_{1},N+2)$ are calculated as follows, see (4,5)\cite{1}
\begin{equation}
E=E_{N_{\alpha }^{ml}}-E_{N_{\alpha }^{ml}N_{\alpha
}^{core}}^{C}+E_{core};E_{1}=E_{N_{\alpha }^{ml+0.5}}-E_{N_{\alpha
}^{ml}N_{\alpha }^{core}}^{C}+E_{core},  \label{2}
\end{equation}
where $E_{N_{\alpha }^{ml}}$ and $E_{N_{\alpha }^{ml+0.5}}$ are the experimental
binding energies of the nuclei with the total number of alpha-clusters equal to
$N_{\alpha }^{ml}$ and $N_{\alpha }^{ml+0.5}$ (for example
$E_{3}=E_{^{12}\mathrm{C}}$ and $E_{3.5}=E_{^{15}\mathrm{N}}$), $ E_{N_{\alpha
}^{ml}N_{\alpha }^{core}}^{C}$ is the energy of the Coulomb interaction between the
peripheral molecule and the core
\begin{equation}
E_{N_{\alpha }^{ml}N_{\alpha }^{core}}^{C}=2N_{\alpha }^{ml}2N_{\alpha
}^{core}e^{2}/R_{\alpha },  \label{3}
\end{equation}
$E_{core}$ is the core binding energy
\begin{equation}
E_{core}=E_{\Delta N}+E_{N_{\alpha }^{core}}+6\epsilon _{\alpha \alpha },
\label{4}
\end{equation}
where the binding energy of excess nn-pairs $E_{\Delta
N}=\sum_{1}^{N_{nn}}E_{i_{nn}}$ was approximated (see (12) in \cite{1}) with
the following formula in dependence on the number of excess nn-pairs $N_{nn}$
\begin{equation}
E_{\Delta N}=(21.93-0.762N_{nn}^{2/3})N_{nn},  \label{5}
\end{equation}
$E_{N_{\alpha }^{core}}$ stands for the binding energy of $N_{\alpha }^{core}
$ of core $\alpha $-clusters
\begin{equation}
E_{N_{\alpha }^{core}}=E_{N_{\alpha }^{core}}^{nuc}-E_{{N_{\alpha }^{core}}
}^{C}+E^{st},  \label{6}
\end{equation}
where $E_{N_{\alpha }^{core}}^{nuc}=N_{\alpha }^{core}\epsilon _{\alpha
}^{nuc}+3(N_{\alpha }^{core}-2)\epsilon _{\alpha \alpha }^{nuc}$, $ E_{N_{\alpha
}^{core}}^{C}=1.848(N_{\alpha }^{core})^{5/3}$; $E^{st}$ is the surface tension
energy
\begin{equation}
E^{st}=(N_{\alpha }+1.7)(N_{\alpha }^{core})^{2/3}.  \label{7}
\end{equation}
The original formula $E^{st}=(N_{\alpha }+1.7)(N_{\alpha }-4)^{2/3}$ (see (17) in
Ref. \cite{1}) was proposed in Ref. \cite{3} as an approximation function to the
factorized sum of the square radii of $N_{\alpha }-4$ clusters for the nuclei with
$Z\geq 30$. The part $(N_{\alpha }-4)^{2/3}$ is changed here for $(N_{\alpha
}^{core})^{2/3}$. In case when $N_{\alpha }^{ml}$=4, eq. (7) and the original one
coincide. So all the formulas are from \cite{1} with one little change in the
formula for the surface tension energy.

For the isotopes of the nuclei with $Z,Z_1 \leq 29$ the binding energy of the
$\alpha$-clusters is known from the energy of the symmetrical nuclei with $Z=N$ (1).
So the binding energy for the isotopes is to be trivial $E =
N\epsilon_{\alpha}+3(N_{\alpha}-2)\epsilon_{\alpha \alpha}+E_{\Delta N}$. The energy
can be written in the terms of core and a peripheral molecule (2) with $E_{core} =
N_\alpha^{core} \epsilon_{\alpha}+3(N_\alpha^{core}-2)\epsilon_{\alpha
\alpha}+6\epsilon_{\alpha \alpha}+E_{\Delta N} +
E^C_{N_\alpha^{ml}N_\alpha^{core}}$. The energy E (2) for the cases almost does not
depend on $N_\alpha^{ml}$ varied within 2 $\div$ 5.

Two molecules on the surface of the core, let them consist of $N_{\alpha }^{ml1}$
and $N_{\alpha }^{ml2}$ alpha-clusters, are needed in describing the isotopes with a
considerable deficiency of excess neutrons in the nuclei with big $Z$, see section
2. Then the binding energy is as follows \cite{1}
\begin{equation}
E=E_{N_{\alpha }^{ml1}}+E_{N_{\alpha }^{ml2}}-(E_{N_{\alpha
}^{ml1}(N_{\alpha }^{core}+N_{\alpha }^{ml2})}^{C}+E_{N_{\alpha
}^{ml2}N_{\alpha }^{core}}^{C})+E_{core}.  \label{8}
\end{equation}
For the odd nucleus $A_{1}(Z_{1},N+2)$ to calculate $E_{1}$ one pn-pair is added to
one of the molecules. Here $E_{N_{\alpha }^{ml1}}$ is exchanged for $E_{N_{\alpha
}^{ml1+0.5}}$. The value $E_{core}$ is calculated by (4) with $N_{\alpha
}^{core}=N_{\alpha }-(N_{\alpha }^{ml1}+N_{\alpha }^{ml2})$. Unlike \cite{1} the
total number of short links in the core does not depend on whether there is one or
two molecules on the surface of the core.

The following formulas (9-12) have been proposed to estimate radii of the nuclei
$A(Z,N)$ on $N_{\alpha }=Z/2$ and the odd nuclei $A_{1}(Z_{1},N+2)$ on $N_{\alpha
}+0.5$ \cite{1,2,3}. The simplest one for $Z,Z_{1}\geq 24$ is the following
\begin{equation}
R=r_{\alpha }N_{\alpha }^{1/3}.  \label{9}
\end{equation}
The core prevails in the cases, so the value $r_{\alpha }$ is to be the charge
radius of a core $\alpha $-cluster. To define $r_{\alpha }$ the empirical radii in
\cite{5}, Table IIIA, with 116 data for the isotopes $A(Z,N)$ and $A_{1}(Z_{1},N+2)$
with $Z\geq 24$ have been fitted. The authors analyze the muonic atom transition
energies with using the same model for all observed there isotopes for the nuclei
with $Z,Z_{1}\leq 60$ and 77 $<Z,Z_{1}\leq 83$. The value $r_{\alpha }=1.595$ fm has
been obtained here with the rms deviation from the experimental data $\delta$ =
0.030 fm. For the 147 isotopes $A(Z,N)$ and $A_{1}(Z_{1},N+2)$ with $Z\geq 6 $ of
the table $\delta$ = 0.064 fm.

The charge radius for the nuclei with $Z,Z_1 \geq$ 6 can be estimated from adding
the volumes of the charges of the core and the peripheral molecule \cite{1}
\begin{equation}
R^{3}=r_{^{4}He}^{3}N_{\alpha }^{ml}+r_{\alpha }^{3}N_{\alpha }^{core},
\label{10}
\end{equation}
where the radius of a peripheral alpha-cluster $r_{^{4}He}$=1.71 fm. For the nuclei
without core, i.e. with $N_{\alpha }\leq 5$, $N_{\alpha }^{core}$=0. Fitting the
data of the Table IIIA \cite{5} for the isotopes with $Z\geq 24$ with the values
$N_{\alpha }^{ml}$, obtained from analysis of binding energies, see section 2,
($N_{\alpha }^{ml}=3$ in most of the cases), gives the value $r_{\alpha }$ = 1.574
fm with $\delta $ = 0.033 fm. As for the isotopes with 12 $\leq Z,Z_{1}\leq $ 23 we
tried $N_{\alpha }^{ml}$= 5, which means the nuclear molecule $^{20}$Ne for the even
isotopes and $^{23}$Na for the odd isotopes. Then for the 147 isotopes with $Z\geq
6$ of the table $\delta$ = 0.040 fm.

Another way to calculate charge radii for the nuclei with $Z,Z_1 \geq$ 6 is the
following \cite{2}
\begin{equation}
N_{\alpha }R^{2}=N_{\alpha }^{core}(r_{\alpha }(N_{\alpha
}^{core})^{1/3})^{2}+N_{\alpha }^{ml}R_{\alpha }^{2},  \label{11}
\end{equation}
where the square core radius and the square distance of the position of the last
$\alpha $-cluster in the system of the center of mass of the core (for the $Z,Z_1
\leq$ 23 the empirical values $R_\alpha$ are used, see (17) in \cite{3}) are added
with their weights to be equal to the square radius of the nucleus weighed with
$N_{\alpha }$. In the cases when $N_{\alpha }^{core}<N_{\alpha }^{ml}$ ($N_{\alpha
}<$ 10 and $N_\alpha^{ml}$ = 5) the formula is rewritten for the center of mass of
the peripheral molecule. Then $N_{\alpha }^{ml}$ and $N_{\alpha }^{core}$ exchange
their places and $r_{\alpha }$ is replaced with $r_{^{4}He}$. The value $r_{\alpha
}$ = 1.595 fm obtained from fitting 116 data for the isotopes with $Z\geq 24$ of
Table IIIA \cite{5} gives $\delta $ = 0.038 fm. For the 147 isotopes with $Z\geq 6$
of the table $\delta $ = 0.040 fm.

The next formula estimates the radii for $Z,Z_1 \geq$ 6 by adding the volumes of the
charge of the peripheral $N_{\alpha }^{ml}$ $ \alpha $-clusters and the body of the
core, consisting of the volume occupied by the bodies of the alpha-clusters and the
volume of the excess nn-pairs \cite{2}
\begin{equation}
R_{m}^{3}=N_{\alpha }^{ml}r_{^{4}He}^{3}+N_{\alpha
}^{core}(4r_{p/n}^{3})+N_{nn}(2r_{n}^{3}),  \label{12}
\end{equation}
where $r_{p/n}$=0.954 fm stands for the radius of the volume which is one nucleon's
share in the volume of an $\alpha $-cluster body, $r_{n}$ =0.796 fm is the radius of
one neutron of a nn-pair. The deviation $\delta $ = 0.028 fm from fitting the data
of the isotopes with $Z\geq 24$ of Table IIIA \cite{5}. For the experimental radii
of the 147 isotopes with $Z\geq 6$ of the table $\delta $ = 0.037 fm.

One should notice here that the radius of a core $\alpha $ - cluster
$4^{1/3}r_{p/n}$ is not its mass radius. It rather defines the space which can't be
occupied by excess neutrons, because there is some structure made of pn-pairs. The
results of fitting experimental radii show that the charge radius of an $ \alpha
$-cluster slowly changes with growing $Z$ (otherwise $\delta$ would not depend of
the group of data selected for fitting). So does the $\alpha$- cluster's body
radius.

The real mass radii of all nucleons considered as elementary bricks of a nucleus
should be equal and they can not change, because the binding energies are too small
in comparison with their masses. The well known formula $R=r_{0}A^{1/3}$ of the
liquid drop model with the fitted value $r_{0}$ =0.95 fm for the isotopes with
$Z\geq 24$ of the Table IIIA \cite{5} gives the considerably bigger deviation
$\delta $ = 0.067 fm and for 147 isotopes with $Z\geq 6$ rms deviation is $\delta $
= 0.12 fm, which is considerably worse than those of the equations (9-12).  This
clearly shows that the distances between nucleons in a nucleus are not related to
their mass radii.

The experimental data of the other isotopes with $60\leq Z,Z_{1}\leq $ 77 are
presented in \cite{5} in a separate table, Table IIIC, because they are treated as
deformed ones. The radii for the isotopes calculated by (9-12) with the $ N_{\alpha
}^{ml}$ obtained in the analysis of experimental binding energies (see section 2)
are in an agreement with the data in the table, although they are always less than
the experimental radii on several hundredths of fm.

\section{Binding energies and radii of the isotopes with $N\geq Z$}

For the nuclei with $Z,Z_{1}\geq 30$ the number $N_{\alpha}^{ml}$ is found from the
analysis of experimental binding energies \cite{6}   with average deviation of
calculated energies from experimental values in 2.5 MeV. Some examples are given in
the table in Appendix. All isotopes with growing the number $\Delta N$ have the
$N_{\alpha }^{core}$ increasing and $N_{\alpha}^{ml}$ correspondingly decreasing. At
$\beta$-stable isotopes $N_{\alpha}^{ml}$ reaches the value 2 for some of the
isotopes with $Z = 38\div 43$ and $N^{ml}$ = 4 for some of the isotopes with $66\leq
Z,Z_{1}\leq 79$. For the other $\beta$-stable isotopes $N_{\alpha }^{ml}$= 3. The
$\beta $-stable isotopes with $ Z,Z_{1}\geq 85$ have $N_\alpha^{ml}=4, 5$. The
calculated binding energies of the isotopes with 12 $\leq \ Z,Z_{1}\leq $ 29 do not
depend much on $ N_{\alpha }^{ml}$ varied within 2 $\div $ 5. So for the $\beta
$-stable isotopes of the nuclei with 24 $\leq \ Z,Z_{1}\leq $ 29 it is suggested
that $N_{\alpha }^{ml}$ equals 3 as it does for the heavier nuclei. For the light
nuclei having core $12 \leq Z,Z_1 \leq 23$ the calculations on (10-12) with the
value $N_\alpha^{ml}$ = 5 give a small deviation $\delta$ from the experimental
radii. Spreading the common rule that with growing $A$ the value $N_\alpha^{ml}$
decreases to 3 for these nuclei makes the deviations smaller.

The specific density of core binding energy is calculated as $\rho=
E_{core}/(N_\alpha^{core} v_\alpha)$ where $E_{core}$ is (4) for the isotopes with
$Z,Z_1 \geq$ 30, $v_\alpha = 4/3\pi r_\alpha^3$ at the charge radius of the core
$\alpha$ - cluster $r_\alpha =1.595$ fm taken from fitting the experimental radii by
(9) and (11). The value $\rho$ grows slowly with the number of excess neutrons and
at $ \beta $-stable isotopes it reaches a saturation value $\rho \approx 2.6\pm 0.1$
$ \mathrm{MeV/fm^{3}}$. For example, the $\beta $-stable isotopes with $Z$ = 30, 60
and 80 have $N^{ml}$= 3 and $\rho =2.48\div 2.67$ $\mathrm{MeV/fm^{3}} $, $2.53\div
2.63$ $\mathrm{MeV/fm^{3}}$ and 2.47 $\div$ 2.55 $\mathrm{MeV/fm^{3}}$
correspondingly. Then the narrow strip of binding energies of $\beta$ - stability is
to be outlined by $E$ (2) with $E_{core} = N_{\alpha }^{core}v_{\alpha }\rho$. With
$N_{\alpha }^{ml}$ =3 (for $Z,Z_1 \leq$ 84) and $\rho =2.55$ $\rm MeV/fm^{3}$ the
binding energy $E$ (2) is a function of one variable $N_{\alpha}$ ($Z$ )
\begin{equation}
E=E_{N_{\alpha }^{ml}}+N_{\alpha}^{core}v_{\alpha }\rho -2N_{\alpha
}^{ml}2N_{\alpha}^{core}e^{2}/R_\alpha, \label{13}
\end{equation}
where $N_{\alpha}^{core}=N_{\alpha}-N_{\alpha}^{ml}$ and
$R_\alpha=2.168(N_{\alpha}-4)^{1/3}$ (for the $Z \leq$ 22 the empirical values
$R_\alpha$ are used, see (17) in \cite{3}). In Fig. 1 the graph of the function (13)
is given in comparison with the experimental values of the lightest and the heaviest
even $Z$ beta-stable isotopes depicting the boundaries of $\beta $-stability.  The
width of the strip is determined by the variation of $\rho$ within 2.5 $\div$ 2.7
$\mathrm{MeV/fm^{3}}$ and by the number $N_\alpha^{core}$.  The physical meaning of
$v_{\alpha }\rho$=43.34 MeV is the core binding energy per one $\alpha$ - cluster.
Then the number of excess neutrons of $\beta$ - stability  is defined by (5) and by
the eq. $E_{\Delta N} = 43.34 N_{\alpha }^{core} - E_{N_{\alpha }^{core}}-6\epsilon
_{\alpha \alpha }$.

\begin{figure}
\begin{center}
\includegraphics*[width=10cm]{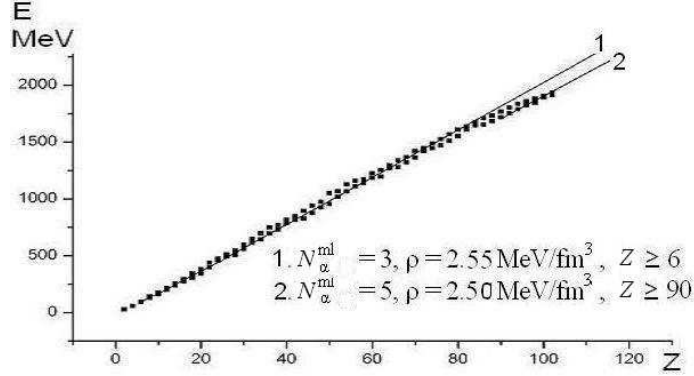}
\end{center}
\caption{The binding energy of beta-stable isotopes. Dots indicate the experimental
energies of the lightest and the heaviest even beta-stable isotopes. Two lines
indicate the function (13) with $N_{\alpha }^{ml}$= 3 and 5.} \label{fig:1}
\end{figure}

As it was shown in \cite{1} the binding energies of the excess nn-pairs in $ \beta
$-stable nuclei obey the eq. (5) with the accuracy in a few MeV. Therefore the
energy $E$ (2,8) with $E_{\Delta N}$ (5) can be calculated only for those isotopes
that have the excess neutrons inside the core, i.e. for the $A\leq A^{st}$ where
$A^{st}$ is the mass number of the heaviest $\beta $-stable isotope among those
belonging to one $Z$. Examples are given in Table 1 (see Appendix) for the isotopes
with $Z$ = 30, 31, 80, 81. For $Z$ = 30 and 31 all the isotopes with $N\geq Z$ with
known energies are given. One can see that for the isotopes with $A > A^{st}$ the
values of the nn-pair separation energy $E_{N_{nn}}^{sep} = E_{A^{\prime
}(Z,N+2)}-E_{A(Z,N)}$ less than $E_{N_{nn}}$ (5). It allows one to suggest that the
nn-pairs come out of the core not affecting the core density. In such cases the
binding energy is equal to the sum of E (2) for the isotope $A^{st}$ and $\sum
E_{i_{nn}}^{sep}$ of the last nn-pairs.

One can see that $E_{N_{nn}}^{sep}$ for both isotopes $A(Z,N)$ and
$A_{1}(Z_{1},N+2)$ in most of the cases are almost equal. It approves the idea that
the cores of the nuclei with $Z$ and $Z_{1}$ are mostly equal. One can see from
Table 1, that the separation energies $E_{N_{nn}}^{sep}$ for the most of the pairs
of $A(Z,N)$ and $A_{1}(Z_{1},N+2)$ differ within 1 MeV. It allows one to make a
prediction for some isotopes with $A>A^{st}$, when the energy of the complementary
nucleus has been already measured. Some of the predicted values are $E_{123(45,78)}$
=1007 MeV, $E_{126(46,80)}$ = 1034 MeV, $E_{139(51,88)}$= 1132 MeV, $E_{145(53,92)}$
= 1171 MeV, $E_{148(54,94)} $ = 1193 MeV, $E_{189(73,116)}$=1500 MeV,
$E_{193(75,118)}$=1529 MeV, $ E_{175(81,94)}$=1348 MeV, $E_{261(101,160)}$=1929 MeV.

In accordance with the representation of a core as a liquid drop one can suggest
that the excess nn-pairs concentrate inside the core near its surface due to the
surface tension, so that the filling out the space inside the core goes from surface
to center. The very first nn-pairs can be at the boundary of the core defined by
core's charge radius. Therefore for the isotopes lighter than stable ones the
equations (9-11) should predict the charge radii. For the $\beta$ - stable isotopes
and the heavier ones with the nn-pairs out of the core one should use (12) to
estimate the nuclear size. Then the nuclear radii with growing $A$ should first
decrease with decreasing $N_{\alpha }^{ml}$ till the stable isotopes, which have
approximately equal radii calculated by (9-12). For the heavier isotopes the radii
slowly increase (12) with growing $A$ because of the growing amount of nn-pairs on
the surface of the core.

\section{Appendix}
\begin{center}
{Table 1. Binding energies and radii of the isotopes with $Z$=30, 31, 80, 81.
$\Delta N$ is the number of excess neutrons in the core. $E_{exp}$ is from \cite{6},
$E^{sep}_{nn}$ is the separation energy of the last nn-pair, $ N_\alpha^{ml}$ is the
number of alpha-clusters out of the core, $R_{exp}$ is from \cite{5}. * indicates a
$\beta$-stable nucleus}
\end{center}
{\begin{tabular}{cccccccccccc} \hline $Z$ & $\Delta N$ & $A$ & $E_{exp}$ &
$E^{sep}_{nn}$ & $N_\alpha^{ml}$ & $E$
(2,8) & $R_{exp}$ & $R$(12) & $R$(11) & $R$(10) & $R$(9) \\
&  &  & MeV & MeV &  & MeV & fm & fm & fm & fm & fm \\ \hline
30 & 0 & 60 & 515 & 0 & 3 & 514 &  & 3.841 & 3.914 & 3.954 & 3.934 \\
31 & 0 & 63 & 541 & 0 & 3 & 538 &  & 3.897 & 3.949 & 4.006 & 3.977 \\
30 & 2 & 62 & 538 & 23 & 3 & 535 &  & 3.864 & 3.914 & 3.954 & 3.934 \\
31 & 2 & 65 & 563 & 22 & 3 & 559 &  & 3.919 & 3.949 & 4.006 & 3.977 \\
\end{tabular}}
{\begin{tabular}{cccccccccccc} \hline $Z$ & $\Delta N$ & $A$ & $E_{exp}$ &
$E^{sep}_{nn}$ & $N_\alpha^{ml}$ & $E$
(2,8) & $R_{exp}$ & $R$(12) & $R$(11) & $R$(10) & $R$(9) \\
&  &  & MeV & MeV &  & MeV & fm & fm & fm & fm & fm \\ \hline
30 & 4 & *64 & 559 & 21 & 3 & 556 & 3.928 & 3.886 & 3.914 & 3.954 & 3.934 \\
31 & 4 & 67 & 583 & 20 & 3 & 579 &  & 3.941 & 3.949 & 4.006 & 3.977 \\
30 & 6 & *66 & 578 & 19 & 3 & 575 & 3.948 & 3.908 & 3.914 & 3.954 & 3.934 \\
31 & 6 & *69 & 602 & 19 & 3 & 599 & 3.996 & 3.962 & 3.949 & 4.006 & 3.977 \\
30 & 8 & *68 & 595 & 17 & 3 & 594 & 3.965 & 3.930 & 3.914 & 3.954 & 3.934 \\
31 & 8 & *71 & 619 & 17 & 3 & 618 & 4.011 & 3.984 & 3.949 & 4.006 & 3.977 \\
30 & 10 & 70 & 610 & 16 & 3 & 613 & 3.983 & 3.952 & 3.914 & 3.954 & 3.934 \\
31 & 10 & 73 & 635 & 16 & 3 & 636 &  & 4.005 & 3.949 & 4.006 & 3.977 \\
30 & 12 & 72 & 626 & 16 & 3 &  &  & 3.973 & 3.914 & 3.954 & 3.934 \\
31 & 12 & 75 & 650 & 16 & 3 &  &  & 4.025 & 3.949 & 4.006 & 3.977 \\
30 & 14 & 74 & 640 & 14 & 3 &  &  & 3.995 & 3.914 & 3.954 & 3.934 \\
31 & 14 & 77 & 663 & 13 & 3 &  &  & 4.046 & 3.949 & 4.006 & 3.977 \\
30 & 16 & 76 & 652 & 13 & 3 &  &  & 4.015 & 3.914 & 3.954 & 3.934 \\
31 & 16 & 79 & 676 & 13 & 3 &  &  & 4.067 & 3.949 & 4.006 & 3.977 \\
30 & 18 & 78 & 663 & 11 & 3 &  &  & 4.036 & 3.914 & 3.954 & 3.934 \\
31 & 18 & 81 & 688 & 12 & 3 &  &  & 4.087 & 3.949 & 4.006 & 3.977 \\
30 & 20 & 80 & 674 & 11 & 3 &  &  & 4.057 & 3.914 & 3.954 & 3.934 \\
31 & 20 & 83 & 695 & 7 & 3 &  &  & 4.107 & 3.949 & 4.006 & 3.977 \\
30 & 22 & 82 & 681 & 7 & 3 &  &  & 4.077 & 3.914 & 3.954 & 3.934 \\
31 & 22 & 85 & 702 & 7 & 3 &  &  & 4.127 & 3.949 & 4.006 & 3.977 \\
80&12& 172&1327&   &3+3 &1327&   &  5.362&   5.512&   5.458&   5.455\\
81&12& 175&    &   &3+3&1349&  &  5.391&   5.536&   5.486&   5.477\\
80&14& 174&1348&21  &3+3 &1343&   &  5.373&   5.512&   5.458&   5.455 \\
81&14& 177&1369&    &3+3 &1367&   &  5.385&   5.522&   5.474&   5.477\\
80&16& 176&1370&22  &2+3 &1369&   &  5.367&   5.498&   5.446&   5.455\\
81&16& 179&1390&21  &2+3 &1392&   &  5.396&   5.522&   5.474&   5.477\\
80&18& 178&1390&20  &2+3 &1386&   &  5.379&   5.498&   5.446&   5.455\\
81&18& 181&1410&20  &2+3 &1409&   &  5.408&   5.522&   5.474&   5.477\\
\end{tabular}}

{\begin{tabular}{cccccccccccc}\hline $Z$&$\Delta
N$&$A$&$E_{exp}$&$E^{sep}_{nn}$&$N_\alpha^{ml}$&$E$(2,8)&$R_{exp}$&$R$(12)&$R$(11)
&$R$(10)&$R$(9)\\
&  &  & MeV & MeV &  & MeV & fm & fm & fm & fm & fm \\\hline
80&20& 180&1410&20  &1+4 &1411&   &  5.391&   5.498&   5.446&   5.455\\
81&20& 183&1430&20  &1+4 &1431&   &  5.419&   5.522&   5.474&   5.477\\
80&22& 182&1430&20  &1+4 &1427&   &  5.402&   5.498&   5.446&   5.455\\
81&22& 185&1450&20  &1+4 &1448&   &  5.431&   5.522&   5.474&   5.477\\
80&24& 184&1449&20  &1+3 &1450&   &  5.396&   5.485&   5.433&   5.455\\
81&24& 187&1468&18  &1+4 &1463&   &  5.425&   5.509&   5.461&   5.477\\
80&26& 186&1467&18  &1+3 &1466&   &  5.408&   5.485&   5.433&   5.455\\
81&26& 189&1487&19  &1+3 &1489&   &  5.436&   5.509&   5.461&   5.477\\
80&28& 188&1485&18  &1+3 &1480&   &  5.419&   5.485&   5.433&   5.455\\
81&28& 191&1504&17  &1+3 &1504&   &  5.447&   5.509&   5.461&   5.477\\
80&30& 190&1502&17  & 4  &1505&   &  5.431&   5.485&   5.433&   5.455 \\
81&30& 193&1526&22  & 4  &1525&   &  5.459&   5.509&   5.461&   5.477\\
80&32& 192&1519&17  & 4  &1519&   &  5.442&   5.485&   5.433&   5.455\\
81&32& 195&1539&13  & 5  &1536&   &  5.487&   5.522&   5.474&   5.477\\
80&34& 194&1535&16  & 4  &1533&   &  5.453&   5.485&   5.433&   5.455\\
81&34& 197&1555&16  & 4  &1553&   &  5.481&   5.509&   5.461&   5.477\\
80&36&*196&1551&16  & 3  &1554&    &  5.448&   5.475&   5.421&   5.455\\
81&36& 199&1571&16  & 4  &1566&    & 5.492&   5.509&   5.461&   5.477\\
80&38&*198&1566&15  & 3  &1567&5.448&  5.459&   5.475&   5.421&   5.455\\
81&38& 201&1586&15  & 3  &1590&    &  5.487&   5.499&   5.449&   5.477\\
80&40&*200&1581&15  & 3  &1580&5.457&  5.470&   5.475&   5.421&   5.455\\
81&40&*203&1601&15  & 3  &1603&5.472&  5.498&   5.499&   5.449&   5.477\\
80&42&*202&1595&14  & 3  &1592&5.467&  5.481&   5.475&   5.421&   5.455\\
81&42&*205&1615&14  & 3  &1615&5.483&  5.509&   5.499&   5.449&   5.477\\
80&44&*204&1609&14  & 3  &1604&5.478&  5.492&   5.475&   5.421&   5.455\\
81&44& 207&1628&13  & 3  &1627&    &  5.520&   5.499&   5.449&   5.477\\\hline
\end{tabular}}

\begin{thebibliography}{00}
\bibitem{1}
G. K. Nie, arXiv:0707.4291v3 [nucl-th] 20Sep2007
\bibitem{2}
G. K. Nie,  {\it Mod. Phys. Lett. A},  {\bf 21}, 1889 (2006).
\bibitem{3}
G. K. Nie,  {\it Mod. Phys. Lett. A}, {\bf 22}, 227 (2007).
\bibitem{4}
P. D. Norman, {\it Eur. J. Phys.} {\bf 14}, 36 (1993).
\bibitem{5}
G. Fricke et al, {\it Atomic  Data and Nuclear Data Tables} {\bf 60}, 207 (1995)
\bibitem{6}
CDFE online service,  http://cdfe.sinp.msu.ru/
\end{thebibliography}
\end{document}